\documentclass[submitting,
			   keeplastbox,
			   nospread,
			   hyphens,
			   titlepage,
			   boxit]{nst}

\usepackage{
	multirow,
	ragged2e,
	subfigure, 
	booktabs,
	xcolor,
	amsmath,
	graphicx,
	geometry,
	amsfonts,
	flushend,
	float,
	url,
	dcolumn,
	epstopdf,			
	upgreek,			
	indentfirst
	}

\usepackage{siunitx}
\usepackage[USenglish]{babel}
\usepackage[utf8]{inputenc}
\usepackage{natbib}

\graphicspath{{figures/}}

\makeatother
\begin{document}

\title{Design and Commissioning of Beam Distribution Line at the SXFEL User Facility}
\thanks{This work was supported by the National Key Research and Development Program of China (2018YFE0103100), the National Natural Science Foundation of China (12125508, 11935020), Program of Shanghai Academic/Technology Research Leader (21XD1404100), and Shanghai Pilot Program for Basic Research – Chinese Academy of Science, Shanghai Branch (JCYJ-SHFY-2021-010).}

\author{Si Chen}
\email{chensi@zjlab.org.cn}
\affiliation{Shanghai Advanced Research Institute, Chinese Academy of Sciences, Shanghai 201210, China}
\affiliation{Shanghai Synchrotron Radiation Facility, Chinese Academy of Sciences, Shanghai 201204, China}
\author{Kaiqing Zhang}
\affiliation{Shanghai Advanced Research Institute, Chinese Academy of Sciences, Shanghai 201210, China}
\affiliation{Shanghai Synchrotron Radiation Facility, Chinese Academy of Sciences, Shanghai 201204, China}
\author{Zheng Qi}
\affiliation{Shanghai Advanced Research Institute, Chinese Academy of Sciences, Shanghai 201210, China}
\affiliation{Shanghai Synchrotron Radiation Facility, Chinese Academy of Sciences, Shanghai 201204, China}
\author{Tao Liu}
\affiliation{Shanghai Advanced Research Institute, Chinese Academy of Sciences, Shanghai 201210, China}
\affiliation{Shanghai Synchrotron Radiation Facility, Chinese Academy of Sciences, Shanghai 201204, China}
\author{Chao Feng}
\affiliation{Shanghai Advanced Research Institute, Chinese Academy of Sciences, Shanghai 201210, China}
\affiliation{Shanghai Synchrotron Radiation Facility, Chinese Academy of Sciences, Shanghai 201204, China}
\author{Haixiao Deng}
\affiliation{Shanghai Advanced Research Institute, Chinese Academy of Sciences, Shanghai 201210, China}
\affiliation{Shanghai Synchrotron Radiation Facility, Chinese Academy of Sciences, Shanghai 201204, China}
\author{Bo Liu}
\affiliation{Shanghai Advanced Research Institute, Chinese Academy of Sciences, Shanghai 201210, China}
\affiliation{Shanghai Synchrotron Radiation Facility, Chinese Academy of Sciences, Shanghai 201204, China}
\author{Zhentang Zhao}
\email[Corresponding author, ]{zhaozhentang@zjlab.org.cn}
\affiliation{Shanghai Advanced Research Institute, Chinese Academy of Sciences, Shanghai 201210, China}
\affiliation{Shanghai Synchrotron Radiation Facility, Chinese Academy of Sciences, Shanghai 201204, China}

\begin{abstract}
As an important measure of improving the efficiency and usability of X-ray free electron laser facilities, simultaneous operation of multiple undulator lines realized by a beam distribution system has become a standard configuration in the recent built XFEL facilities. In Shanghai, SXFEL-UF, the first soft X-ray free electron laser user facility in China, has finished construction and started commissioning recently. Electron beam from linac is alternately distributed between the two parallel undulator beam lines by a beam distribution system with a \SI{6}{\degree} deflection line. The beam distribution system is designed to keep the beam properties like low emittance, high peak charge and small bunch length from being spoiled. Beam collective effects such as the dispersion, coherent synchrotron radiation and micro-bunching instability should be well suppressed to guarantee the beam quality. In this work, the detailed physics design of the beam distribution system is described and the recent commissioning result is reported.

\end{abstract}


\maketitle
\nolinenumbers
\section{Introduction}\label{sec.I}

In the recent few decades, with the rapid development of high-gain Free Electron Lasers (FELs), X-ray FEL facilities based on the high-performance RF linear accelerator, have become a powerful platform for extensive research areas including but not limited to the bioscience, material science, chemistry, etc.\cite{Huang2021}. To satisfy the growing thirsty of various users in various communities, several X-ray Free Electron Laser user facilities have been built world wide, including several hard X-ray FEL facilities such as the LCLS\cite{Emma2010}, SACLA\cite{Ishikawa2012}, PAL-FEL\cite{Kang2017}, European XFEL\cite{Decking2020}, and Swiss-FEL\cite{Prat2020}, and several soft X-ray FEL facilities such as the FLASH\cite{Ackermann2007} and Fermi-FEL\cite{Allaria2012}. In China, several X-ray FEL facilities have been built or under construction, including the Shanghai Soft X-ray FEL Facility (SXFEL)\cite{Zhao2017,Liu2022} and the High repetition rate hard X-ray FEL facility, which is named as SHINE\cite{Zhao2017shine}.

With the rapidly growing number of users and their various experiment requirement, it naturally arises a demond of building more XFEL facilities with various parameters of light. However, unlike the storage-ring based sychrotron light sources that can run tens of beamlines simultaneously, linac based XFEL facilities are typically single-user facility in the early stage. Meanwhile, the number of XFEL facilities is limited by the high cost of building one. As a more efficient and economical approach, in the recent built XFEL facilities, the beam distribution system is becoming a standard configuration in order to accommodate multiple undulator lines with various X-ray parameters for feeding more beamlines\cite{Hara2016,Hara2018,Beukers2019,Balandin2011,Milas2012,DiMitri2013}. In such a beam distribution system, electrons from linear accelerator are separated by switching magnets and delivered alternately to each undulator line on a predetermined model, either pulsed or bunch-by-bunch. At the mean time, the beam quality should be well maintained during the beam distribution process, especially when the beam current is high, which is usualy a necessary for high-gain XFEL.

As the first X-ray FEL facility in China, the SXFEL user facility is also designed to be a multi-beamline facility. Two undulator lines are fed by a single linac with 50 Hz bunch repetition rate and more undulator lines can be accommodated in the future. In this work, the detailed design of a beam distribution system to realize multi-beamline operation of the SXFEL-UF is described. The commissioning of the beam distribution system has started in late 2021 and the main results are also presented in this work.

\section{The SXFEL Facility}

\begin{figure*}[t]
	\centering
	\includegraphics*[width=1.0\linewidth]{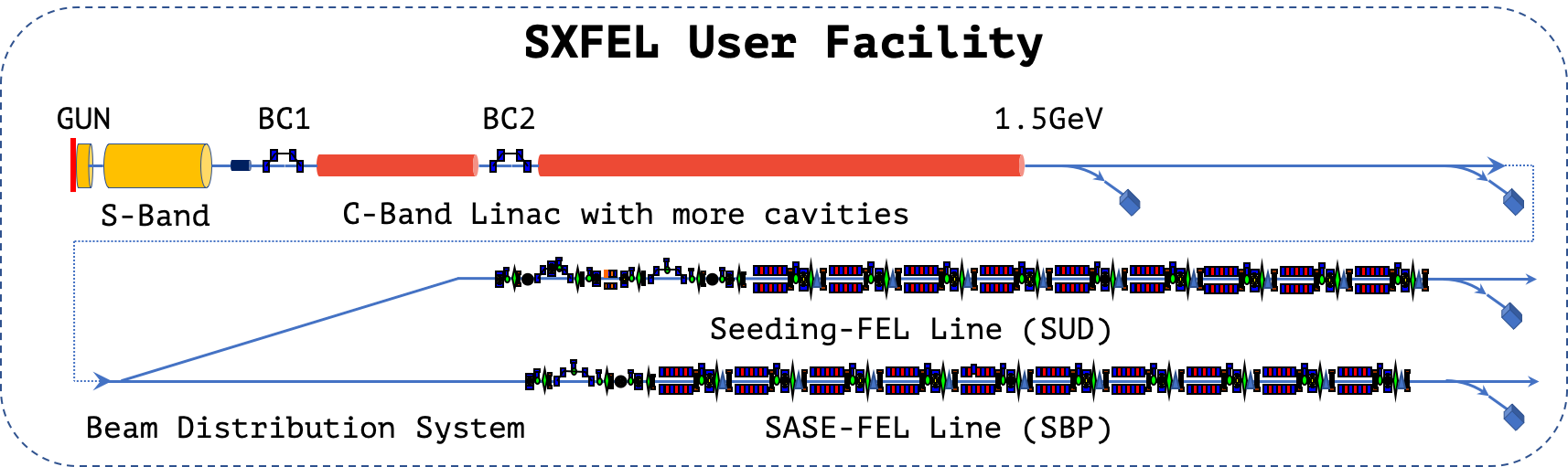}
	\caption{\small{Schematic view of the SXFEL-UF.}}
	\label{fig:tf2uf}
\end{figure*}

The SXFEL facility is aimed to open up the enormous field related to XFEL in China and to accumulate the indispensable technical experience for constructing and utilizing the future hard x-ray FEL facility. It is developed in phases. First a test facility (SXFEL-TF) with a \SI{840}{MeV} linac is built for generation of \SI{8.8}{\nano\meter} full coherent soft X-ray radiation and demonstration of various seeded FEL mechanisms. In 2020, it has achieved its goal with the demonstration of two-stage HGHG-HGHG cascade and two-stage EEHG-HGHG cascade schemes. 

Soon afterwards, the SXFEL-TF has been upgraded and integrated to the user facility (SXFEL-UF)\cite{Liu2022}. It is designed to cover the whole water window range. To accomplish this, the beam energy is upgraded to about \SI{1.5}{GeV} by adding more C-band RF structures to the linac. Two individual undulator lines are parallelly installed in the newly built undulator hall. Directly downstream of the linac it is a brand new SASE-FEL line with radiation wavelength about \SI{2}{\nano \meter}. The existing seeded-FEL line of SXFEL-TF is moved to about \SI{3}{\meter} aside of the SASE-FEL line with an upgrade of more undulator sections for a radiation wavelength about \SI{3}{\nano \meter}. The schematic layout of the SXFEL-UF is shown in Fig. \ref{fig:tf2uf}. Some main beam parameters of SXFEL-UF are shown in Table \ref{table:params}. 

\begin{table}[!hbt]
	\centering
	\caption{Main Parameters of SXFEL Linac}
	\label{table:params}
	\setlength{\tabcolsep}{6mm}{
	\begin{tabular}{c c c}
		\toprule
		\textbf{Parameters} 		& \textbf{Values} 		& \textbf{Units}		\\
		\midrule
		\textit{E} 					& 1.5$\sim$1.6        		& \si{\giga eV}			\\
		\midrule
		$\sigma_E$/$E$ (rms)   		& $\leq$\SI{0.1}{\percent} 	&						\\
		\midrule
		$\varepsilon_{n}$ (rms) 	& $\leq$1.5         		& mm$\cdot$mrad			\\
		\midrule
		$l_{b}$ (FWHM) 				& $\leq$0.7 				& \si{\pico\second}  	\\
		\midrule                                    			
		$Q$ 						& 500 						& \si{\pico\coulomb}  	\\
		\midrule                                    			
		${I_{pk}}$ 					& $\geq$700 				& \si{\ampere}  		\\
		\midrule
		${f_{rep}}$ 				& 50 						& \si{\hertz}  			\\
		\bottomrule
	\end{tabular}}
\end{table}

For the parallel operation of the two FEL lines, a beam distribution system is located between the linac and undulator section. The \SI{50}{hertz} electron bunch train from linac is distributed in two directions, either to the SASE-FEL line or to the seeding-FEL line. Because of the high requirement of the seeded FEL, the beam distribution system should be able to guarantee a stable, precise transportation of the electron beam with a well maintained beam quality properties, such as the low emittance, high peak charge and small bunch length. The detailed physics design of the switchyard is described below.

\section{Beam Distribution System Design of SXFEL-UF}

\subsection{General Layout}

\begin{figure*}[!ht]
	\centering
	\includegraphics*[width=1.0\linewidth]{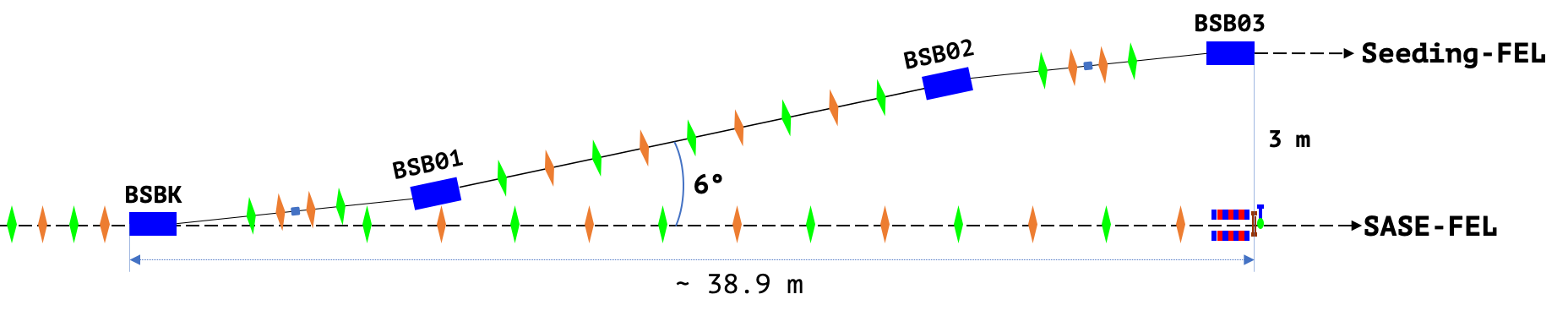}
	\caption{\small{Schematic layouts of the beam distribution system of SXFEL-UF. The total bending angle of the beam distribution dog-leg is about \SI{6}{\degree} with the limitation of total available space for beam distribution system. A pulsed kicker magnet acts as the first bending magnet of the dog-leg with the ability of bunch-by-bunch distribution between the two FEL lines.} }
	\label{fig:dog-leg}
\end{figure*}

The beam distribution system uses a fast switching kicker magnet for bunch-by-bunch distribution of the \SI{50}{hertz} electron bunch train from the linac. It is also designed to be programmable for an arbitrary separation model. When the kicker magnet is on, the electron bunch is deflected horizontally to the seeding-FEL line through a deflection line, otherwise it goes straight to the SASE-FEL line downstream of the linac without any deflection. 

Since the two FEL lines lie parallel in the undulator hall, the deflection line uses a dog-leg structure to bring the kicked beam to the entrance of the seeding-FEL line. Due to the limitation of the longitudinal distance, the total deflection angle of the dog-leg is about \SI{6}{\degree}. The most immediate effect of the dog-leg is the dispersion function. When electron bunch with non-zero energy spread passes through a bending magnet, dispersion effect introduces a coupling between the transverse position and the energy of each electron in the bunch so that the electrons spread transversely afterwards. The transverse phase space can be severely destroyed by this effect. In order to cancel the dispersion of beam distribution dog-leg, its entrance and exit bending magnets are replaced by two identical double-bend-achromat (DBA) sections, respectively. With this configuration, both the dispersion element ($R_{16}$) and the dispersion prime element ($R_{26}$) are cancelled locally after each DBA and globally after the whole dog-leg. The dispersion evolution along the dog-leg is shown in Fig. \ref{fig:optics} as the dashed lines. Between the two DBA sections, several quadruples are inserted for beam matching. The position of the elements is adjusted carefully to avoid conflict between the straight line and the deflected line. The total projected length of the dog-leg is about \SI{39}{\meter}. A schematic view of such a dog-leg is shown in Fig. \ref{fig:dog-leg}.

\subsection{Optics for CSR \& MBI  Suppression}

For high-gain XFEL facilities, typically a high peak current, low emittance electron beam is necessary for obtaining as higher gain in as shorter gain length. However, when such a electron beam passes through the deflection line of the beam distribution system, several kinds beam collective effect may spoil both the transverse and longitudinal phase space of the electron beam, thus reduces the performance of XFEL. Those collective effects should be carefully suppressed in the deflection dog-leg of the beam distribution system. 

\begin{figure}[htb]
	\centering
	\includegraphics*[width=1.0\linewidth]{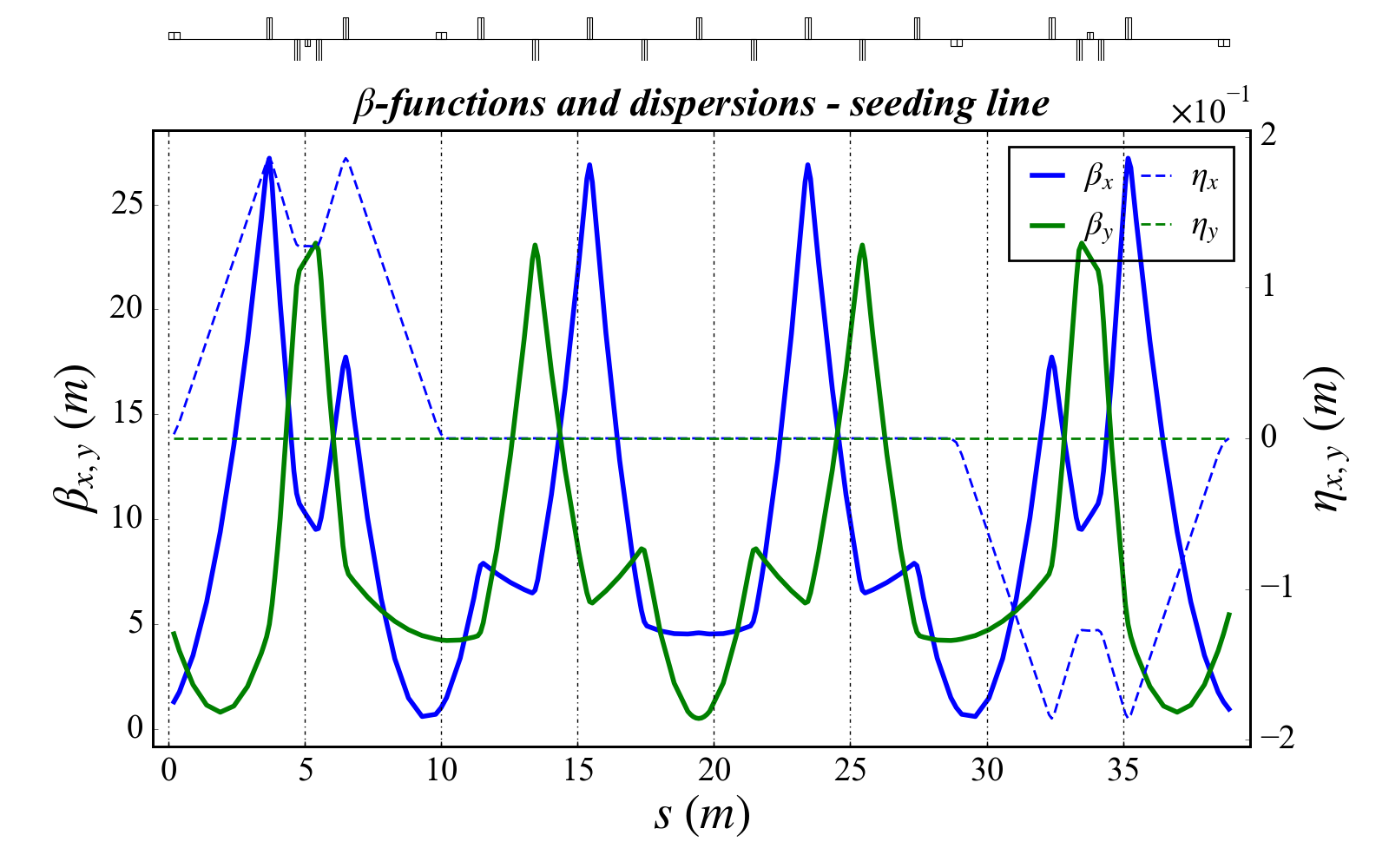}
	\caption{\small{Twiss functions ($\beta_{x,y}$) and dispersion functions ($\eta_{x,y}$) evolution along the dog-leg. The $\beta_x$ at the entrance kicker magnet is optimize to be around 1.6 m so that all the bending magnets have the similar value with the symmetrical optics. The maximum values of $\beta$ and $|\eta|$ are also optimized for smaller beam stay-clear area. }}
	\label{fig:optics}
\end{figure}

Coherent synchrotron radiation (CSR) introduced emittance growth is one of the critical beam dynamic issues of the beam distribution dog-leg. When the electron bunch passes through the bending magnet, synchrotron radiation is emitted. If the bunch length is short enough that it is comparable with the radiation spectral components, the synchrotron radiation becomes coherent. CSR from the bunch tail part may catch up with the head part and interact with the electrons inside as the bunch goes by. Both the transverse and longitudinal phase space distribution can be changed by this process. In longitudinal direction, the CSR field introduces a energy modulation along the longitudinal coordinate of the bunch so that the energy spread increases. In transverse direction, the CSR field mainly acts as a special term of dispersion due to the longitudinal energy variation ans thus the emittance in the bending direction grows. To avoid CSR induced emittance growth in the distribution dog-leg of SXFEL-UF, it is necessary to cancel the transverse effect of CSR. For this purpose, the so called "optics balance method is used. The lattice of the dog-leg is designed to be mirror symmetrical with a small beam waist at each bending magnet to keep the CSR kick small but the same at each bending magnet. Meanwhile, the betatron phase advance between the two DBA sections is matched to be an odd multiple of $\pi$ so that both the CSR kick and the CSR dispersion are cancelled at the end of the distribution dog-leg. The lattice functions of such a dog-leg is shown in Fig. \ref{fig:optics} as the solid lines.

\begin{figure}[ht]
	\centering
	\includegraphics*[width=1.0\linewidth]{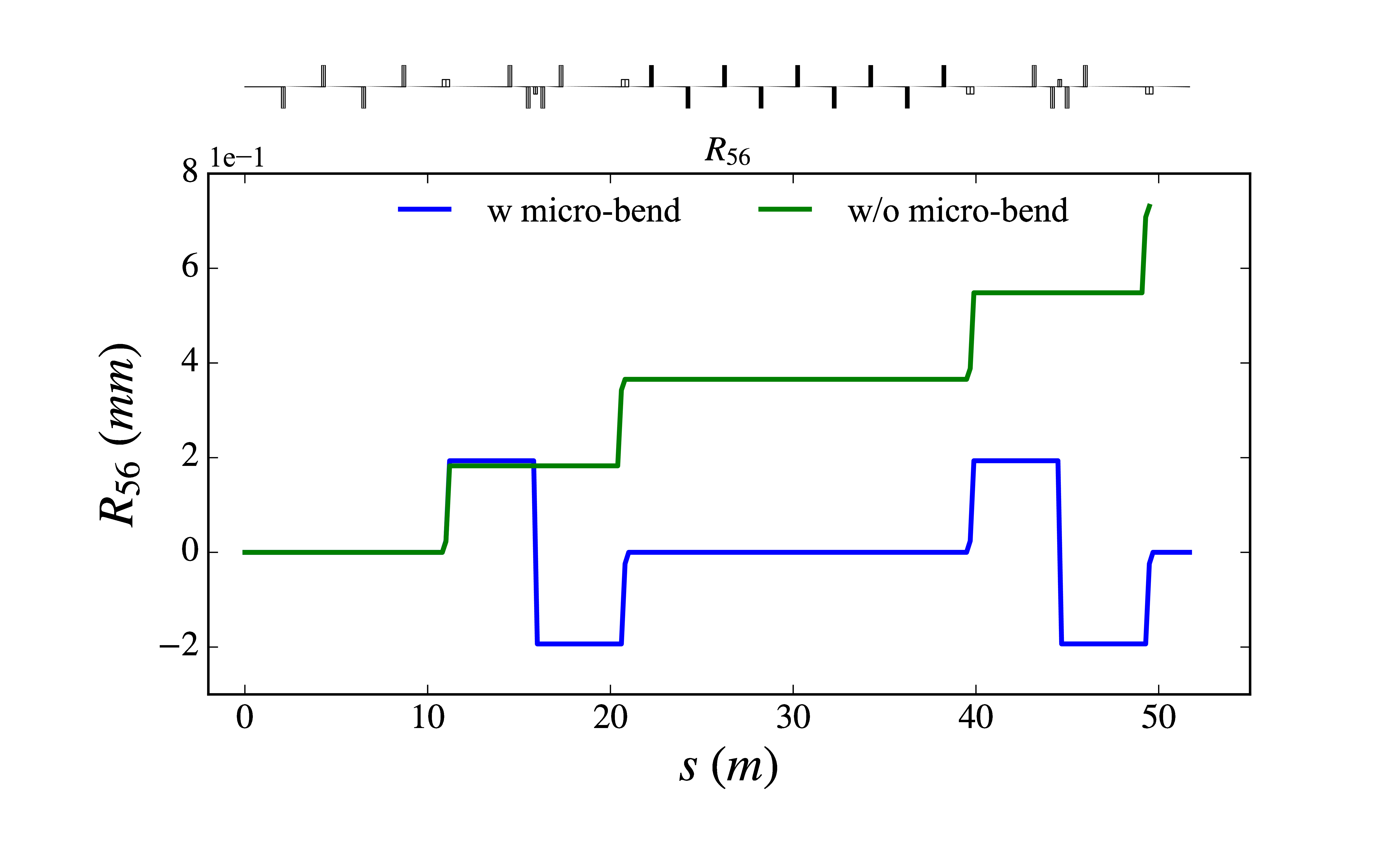}
	\caption{\small{$R_{56}$ evolution along the dog-leg. Without micro-bend, the global $R_{56}$ is over 700 $\mu m$ while with the micro-bend, the $R_{56}$ is eliminated to less than 1 $\mu m$, which is negligible for micro-bunching growth.}}
	\label{fig:r56}
\end{figure}

Another critical beam dynamic issue of the beam distribution system is the micro-bunching instability (MBI). It results from an interplay of various collective effects such as longitudinal space charge (LSC) effect, coherent synchrotron radiation (CSR) in dipole magnets, and the energy-dispersion correlation in magnetic bunch length compressors. An energy modulation is introduced in the beam longitudinal phase space and it is easily being converted to a density modulation while passing through a dispersive magnetic optics section with non-zero $R_{56}$. The amplified density modulation further drives even larger energy and density modulations downstream, thus the beam quality is significantly downgraded. Micro-bunching gain should be well suppressed in the beam distribution system with multi-bend deflection line to guarantee a high spectral brilliance, especially at output radiation wavelengths in EUV and soft x-ray range. For this purpose, a small bending magnet (micro-bend) is inserted in the middle of the DBA cell with a small angle reverse to the DBA deflection angle. With this design, the $R_{56}$ of each DBA becomes zero. Apply this design to both of the DBA cells of the switchyard, it becomes globally isochronous, as is shown in Fig. \ref{fig:r56}. With the isochronous configuration, the deflection line of the beam distribution system is substantially transparent to any incoming modulation induced by micro-bunching instability in the linac.

\subsection{S2E Tracking Results}

\begin{figure}[htb]
	\centering
	\includegraphics*[width=1.0\linewidth]{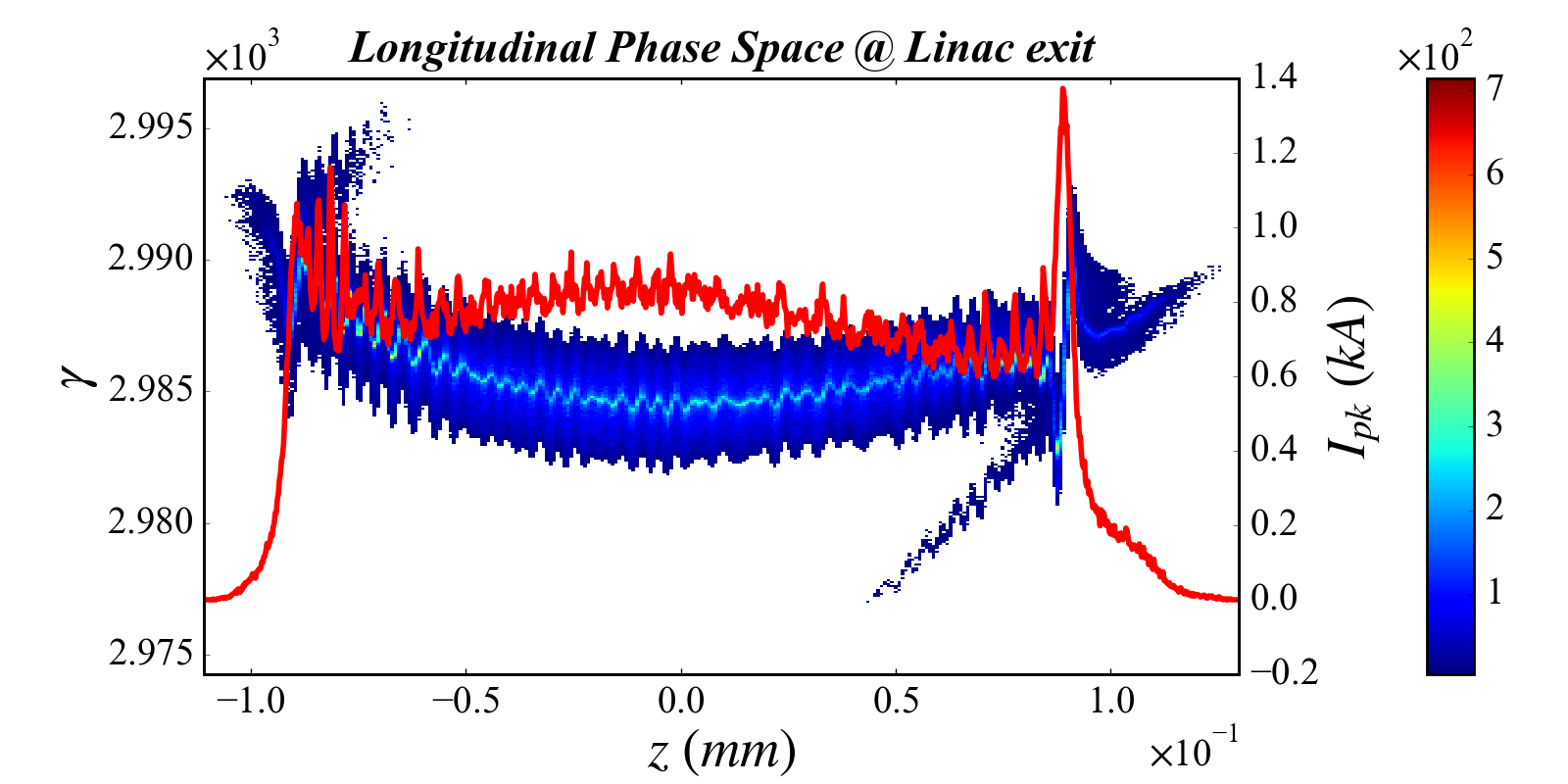}
	\caption{\small{Longitudinal phase space at the linac exit. Beam central energy is about 1.5 GeV, bunch length (FWHM) is about 700 fs, peak current is about 800 A on the flat top and the normalized emittance is about 1.0 mm$\cdot$ mrad.}}	
	\label{fig:longps}
\end{figure}

The start-to-end tracking from the linac end throughout the beam distribution section is performed by the code {\textsc{Elegant}}\cite{elegant}. The longitudinal phase space at the linac exit is shown in Fig. \ref{fig:longps}. Two stage compression bring the 500 pC electron bunch from 10 ps after the injector to less than 0.7 ps at the exit of linac with a long flat top in the longitudinal phase space, which is necessary for multi-stage energy modulation in some complex seeding-FEL mechanisms. Laser heater is used for smoothing the longitudinal phase space and a slice energy spread of about $1\times10^{-4}$ is obtained at last. The peak current on the flat top part is about 800 A and the normalized emittance is about 1.0 $mm\cdot mrad$. 

Fig. \ref{fig:longps} shows the normalized emittance growth ($\varepsilon_{x,f}/\varepsilon_{x,o}$) with respect to the betatron phase advance ($\phi$) between the two DBA cells, where the $f$ and $o$ denoted in the subscripts indicate the values at the end of linac and the beam distribution dog-leg respectively. It is clear that the emittance growth is almost completely eliminated with the $\pi$ phase advance, which indicates that the current optics well satisfies the requirement of suppressing the CSR introduced emittance growth. 

\begin{figure}[!htb]
	\centering
	\includegraphics*[width=1.0\linewidth]{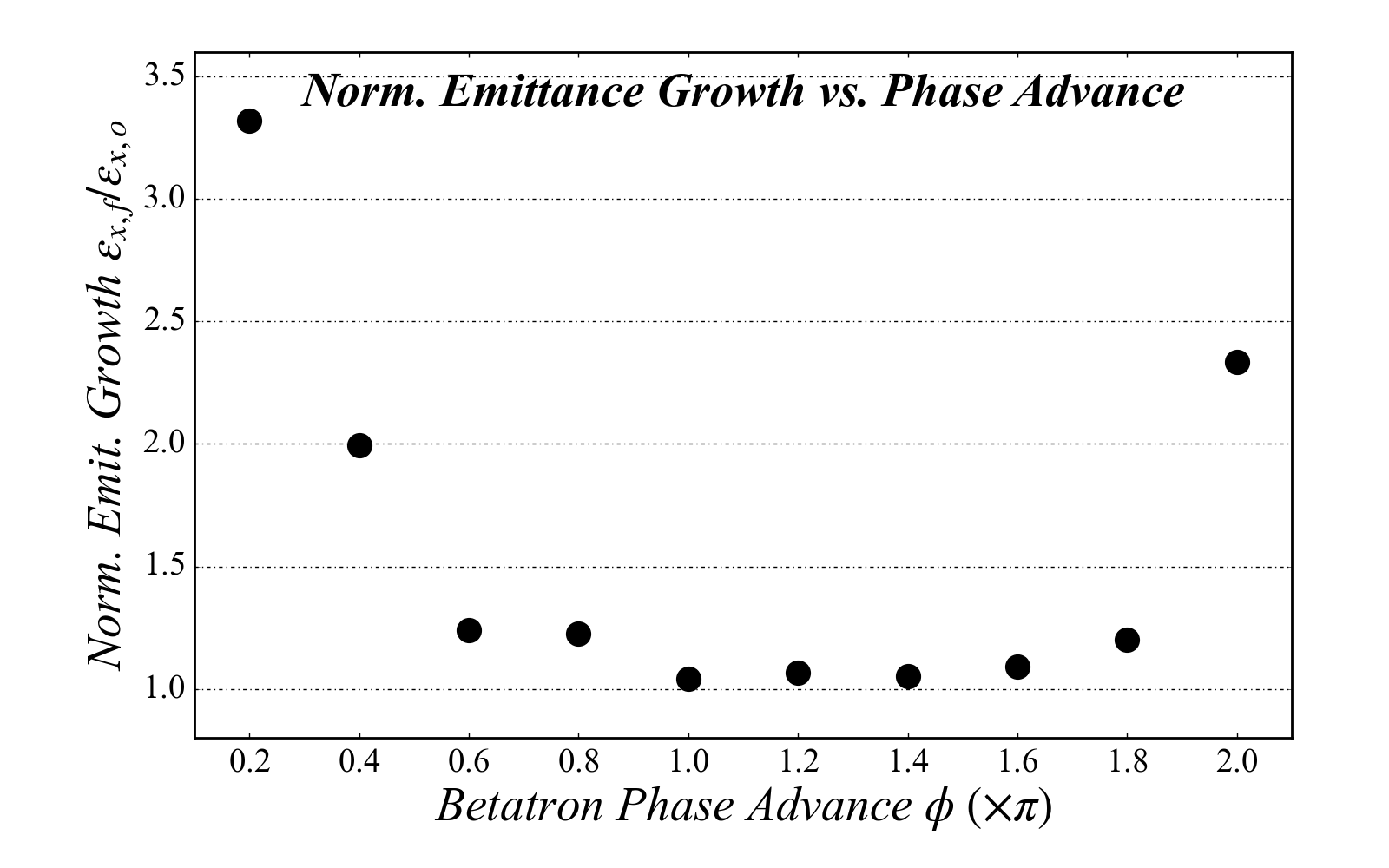}
	\caption{\small{Normalized emittance growth with respect to the betatron phase advance between the two DBA cells. Minimum emittance growth appears at $\phi\sim\pi$, which is much less than the expected \SI{10}{\percent}.}}
	\label{fig:emitt}
\end{figure}

\begin{figure}[!htb]
	\centering
	\includegraphics[width=1.0\linewidth]{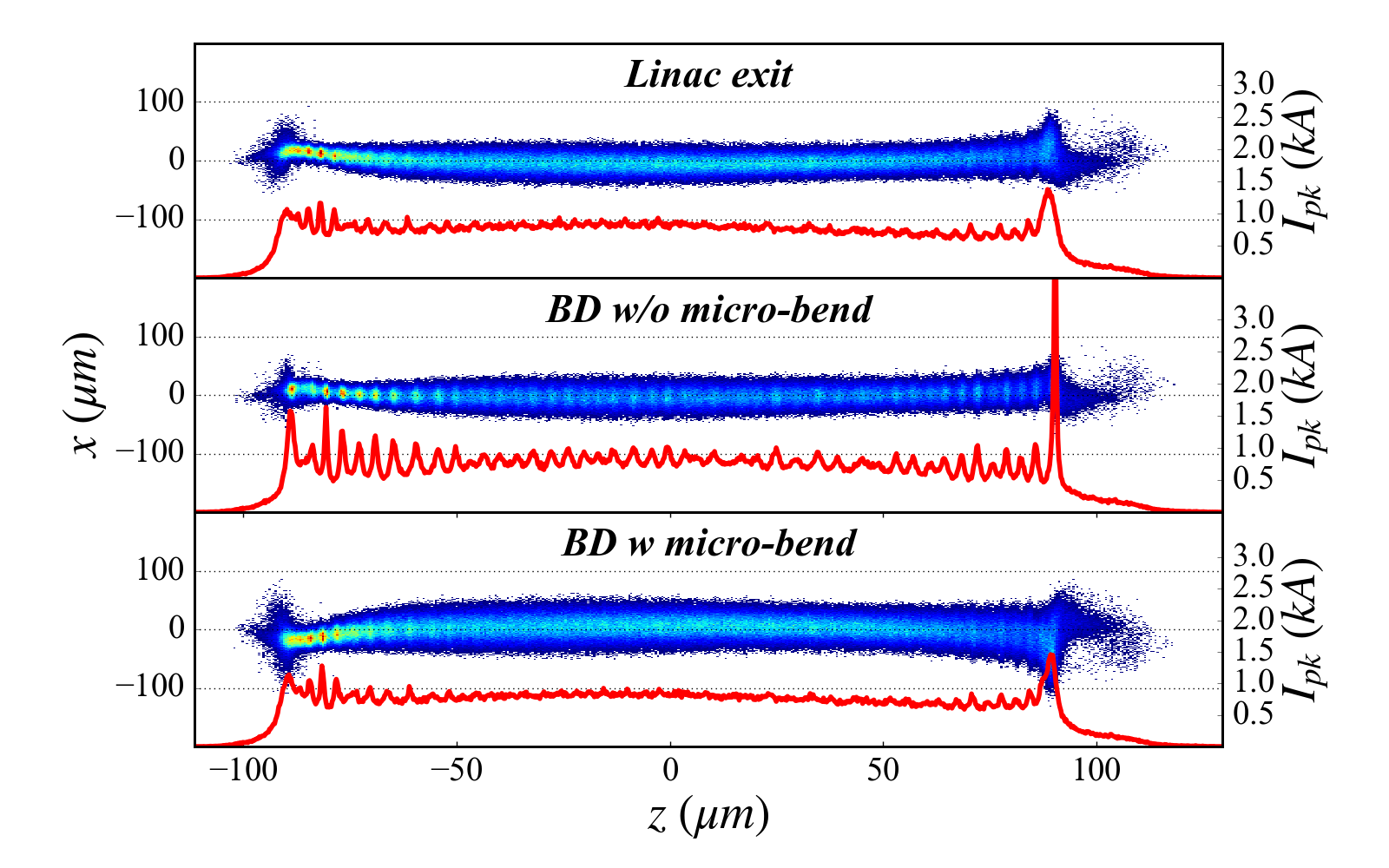}
	\caption{\small{Comparison of the t-x phase space and the current profile at linac exit (upper), BD exit without micro-bend (middle) and BD exit with micro-bend (lower). }}
	\label{fig:comp}
\end{figure}

As is seen in Fig. \ref{fig:longps}, although the laser-heater is used, some residual micro-bunching structures still appear in the longitudinal phase space. Besides, there is a horn with peak current over 1.5 kA in the bunch head part. A comparison of the t-x phase space and current profile before and after the switchyard is shown in Fig. \ref{fig:comp}. For the case that $R_{56}\neq$0, it shows an obvious growth of the microbunching structure in the longitudinal phase space, especially on the head horn part. For the isochronous case with micro-bend, only imperceptible microbunching gain can be observed. The longitudinal phase space is well preserved after the distribution dog-leg. 

\subsection{Trajectory Jitter Analysis}

For sufficient and stable interaction between beam and seed laser, the seeded FEL line requires a transverse beam position jitter less than $0.1\sigma_x$. The major sources of the horizontal trajectory jitter comes from the magnet power fluctuation especially the kicker magnet power jitter. The vertical jitter mainly comes from the quadrupole misalignment jitter due to ground vibration. 

\begin{figure}[htb]
	\centering
	\includegraphics*[width=1.0\linewidth]{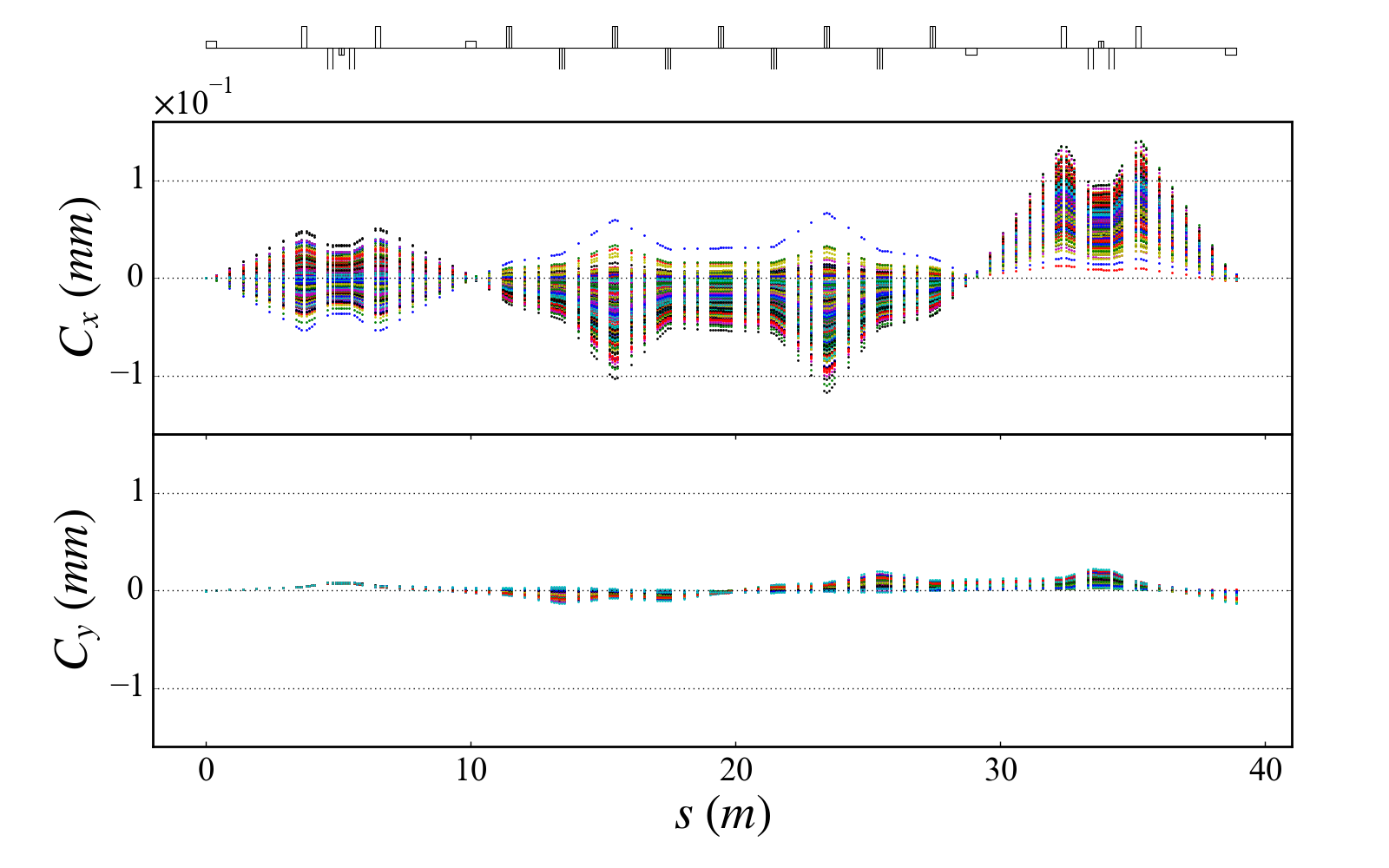}
	\caption{Transverse trajectory jitter of 200 cases of jitter source. The horizontal trajectory jitter is less than \SI{10}{\percent} with respect to the beam size $\sigma_x$ and \SI{<1}{\percent} in vertical direction}
	\label{fig:jitter}
\end{figure}

A simulation of the trajectory jitter of the beam distribution system is done with the kicker power jitter $\sim$ 100 ppm, bending magnet power jitter $\sim$ 50 ppm and ground vibration amplitude $\sim$ \SI{200}{\nano\meter} in both horizontal and vertical direction (all in RMS). Fig. \ref{fig:jitter} shows the transverse trajectory jitter along the beam distribution dog-leg of 200 random seeds of the jitter source. The horizontal position jitter (RMS) at the exit of dog-leg is less than $0.1\sigma_x$ and the vertical position jitter (RMS) is less than \SI{1}{\percent} of $\sigma_y$. However, a further simulation shows that the trajectory jitter is dominated by the kicker jitter and grows almost linearly with it. In a word, 100 ppm is the criteria amplitude of the acceptable kicker power jitter.

\section{Commissioning Results}

The beam distribution system of SXFEL-UF has been installed in the front part of the newly built undulator hall in late 2020. The commissioning of the Switchyard and Seeding FEL line has started at the beginning of November 2021. However, since the power supply of the kicker magnet hasn't reach the expected stability requirement, as is described in the upper section, it is not acceptable for stable operation of seeding-FEL line. So the kicker magnet is absent in this stage of commissioning until it is stable enough and it's function is instead of a DC bending magnet temporary.

The dispersion function measurement is based on measuring the orbit change at each cavity BPM with different beam energy. As is seen in Eq. \ref{eq:disp}, if the dispersion is closed, the orbit data will not change with the energy change, otherwise there will be an orbit vs. relative energy change slope observed. By fitting the slope the dispersion function and even higher order dispersion terms can be obtained. 

\begin{equation}
	x=x_0+R_{16}\frac{\Delta{E}}{E}+T_{166}(\frac{\Delta{E}}{E})^2+...
	\label{eq:disp}
\end{equation}

\begin{figure}[!htb]
	\centering
	\includegraphics*[width=1.0\linewidth]{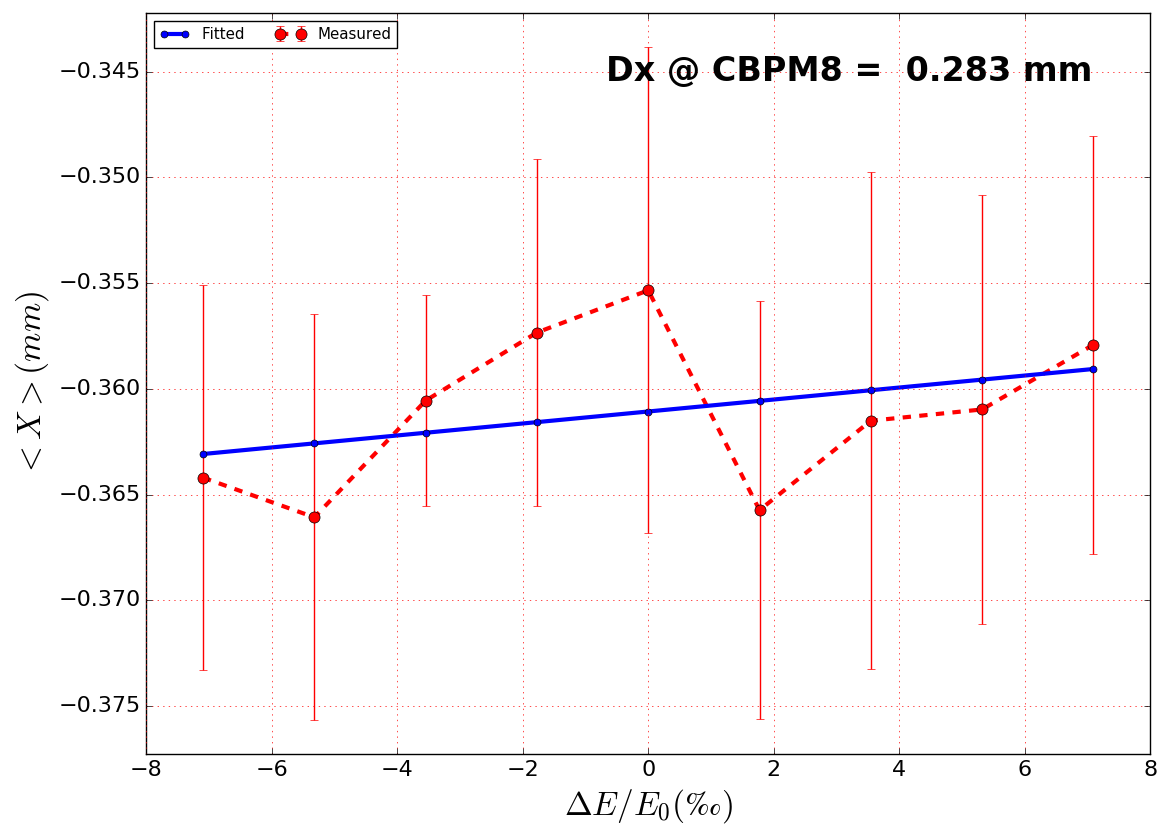}
	\caption{\small{The transverse dispersion is measured by monitoring the position variation on CBPM while varying the bending angle as an equivalence of varying beam energy. }}
	\label{fig:dx-measured}
\end{figure}

\begin{figure}[!htb]
	\centering
	\includegraphics*[width=1.0\linewidth]{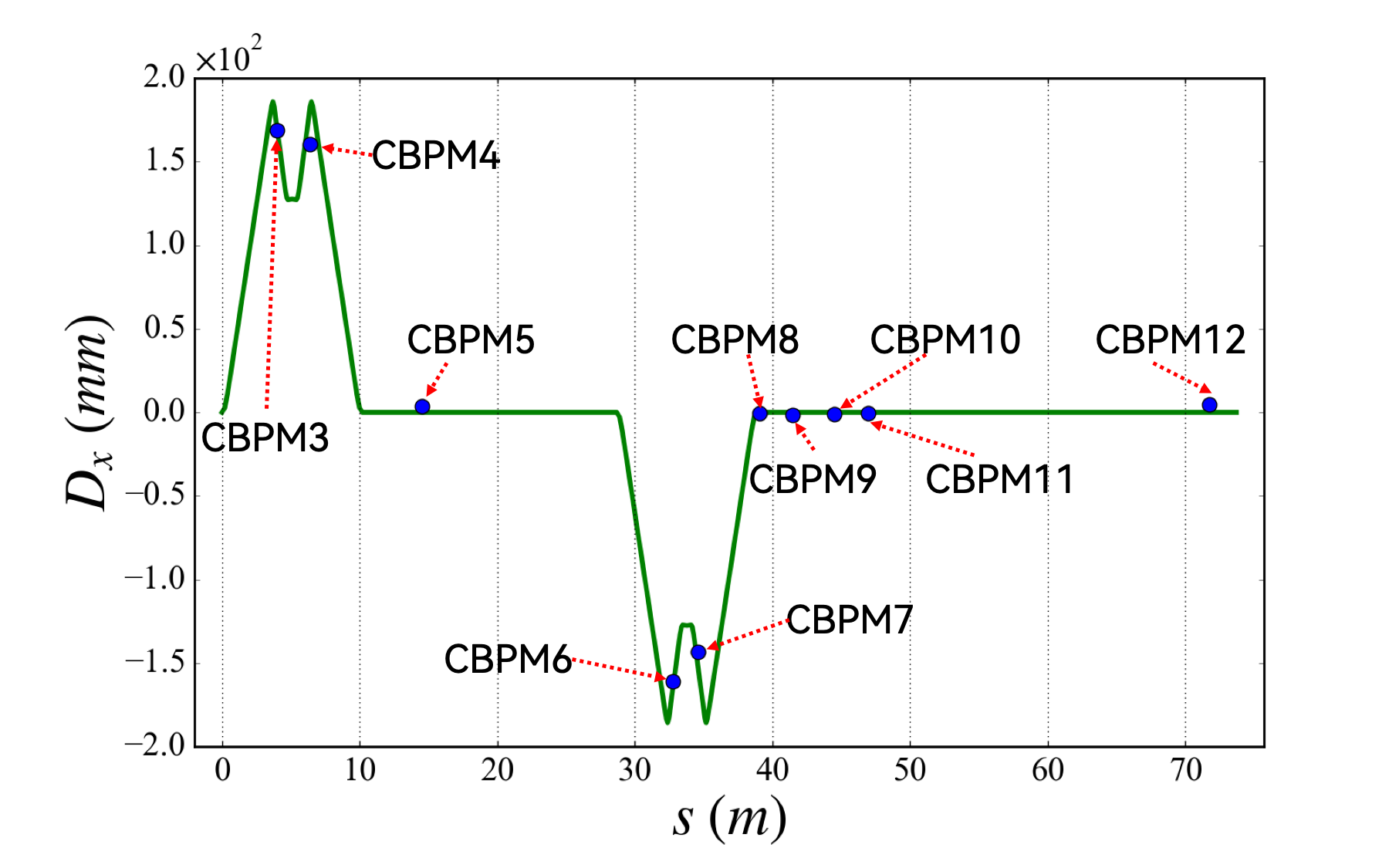}
	\caption{The dispersion is measured on all the CBPMs along and downstream the beam distribution line and the data shows a good consistency with the theoretical value. }
	\label{fig:dx-all}
\end{figure}

Fig. \ref{fig:dx-measured} shows the measurement result of  horizontal dispersion at the exit of distribution dog-leg. The residual horizontal dispersion after the dog-leg is cancelled to less than \SI{1}{\milli\meter} which is much smaller than the required \SI{10}{\milli\meter} value. In order to make sure the $\eta_x^\prime$ is also also well cancelled, dispersion is measured at more CBPMs downstream. Fig. \ref{fig:dx-all} shows the measured dispersion at all the CBPMs from the beam distribution system to the entrance of the FEL line as a comparison with the theoretical value. The result not only confirms the reliability of measurement but also confirms a well cancellation of $\eta_x^\prime$.

\begin{figure}[!htb]
	\centering
	\includegraphics*[width=0.9\linewidth]{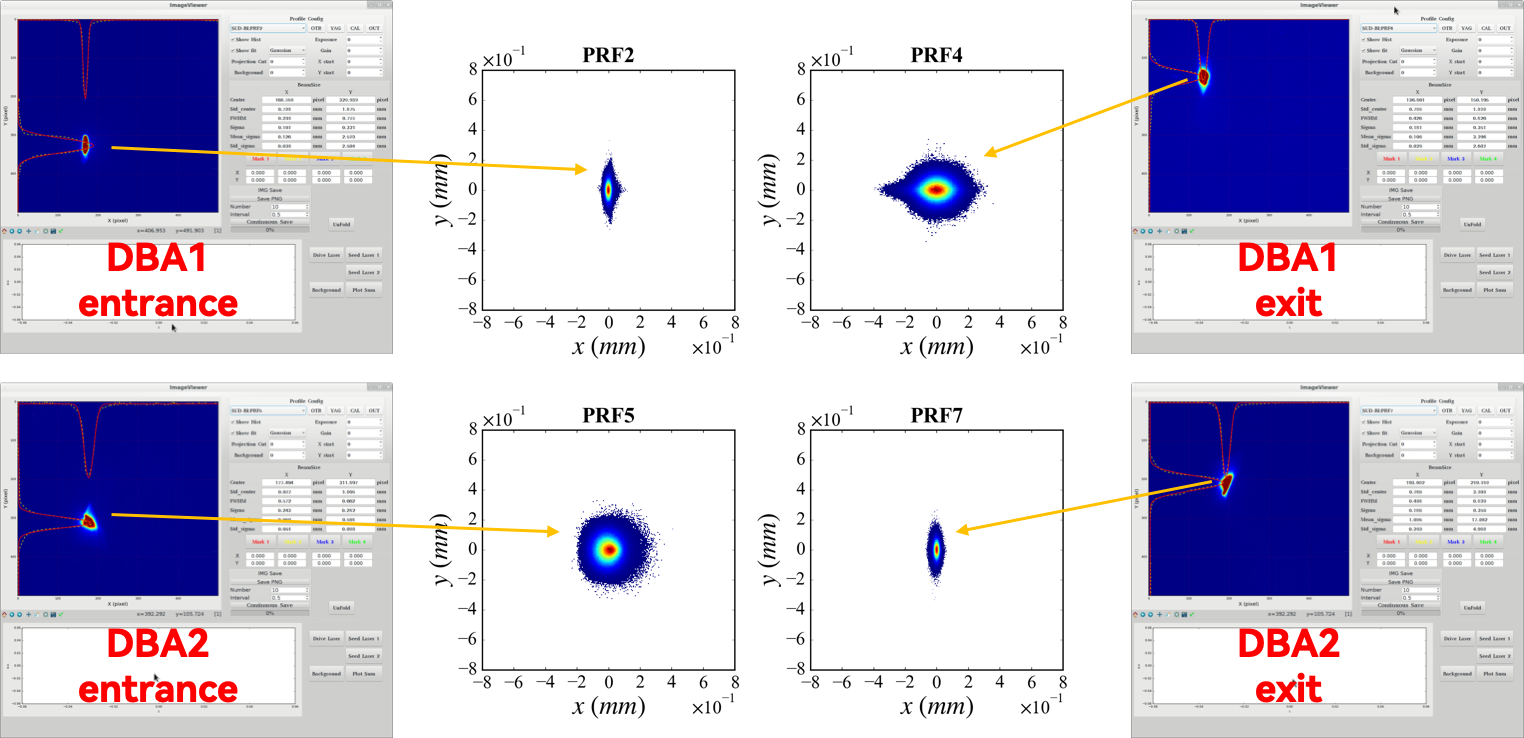}
	\caption{\small{The beam spot at each beam profile monitor as a comparison with the theoretical beam spot. With the well cancelled CSR induced emittance growth, the beam envelop is well sustained along the beam distribution line.}}
	\label{fig:matching}
\end{figure}

The betatron matching of beam distribution system is done by keeping the theoretical configuration of magnets while matching the entrance parameter from the linac. The emittance and twiss parameters are measured by varying the quadrupole and fitting the beam spot variation on a downstream OTR screen. Then the beam is matched from linac exit to the entrance of beam distribution system by an automatic algorism based on the code \textit{Ocelot}. Fig. \ref{fig:matching} shows the comparison between the observed beam spot on each screen and the theoretical beam spot after matching, which shows a well agreement. With such an optics, the emittance is measured after the dog-leg. In order to reduce the fluctuation of emittance measurement, an average of multi-times measurement has been down and the result shows a emittance growth only about \SI{3}{\percent}, which is well below the requested \SI{10}{\percent} emittance growth.

\begin{figure}[!htb]
	\centering
	\includegraphics*[width=1.0\linewidth]{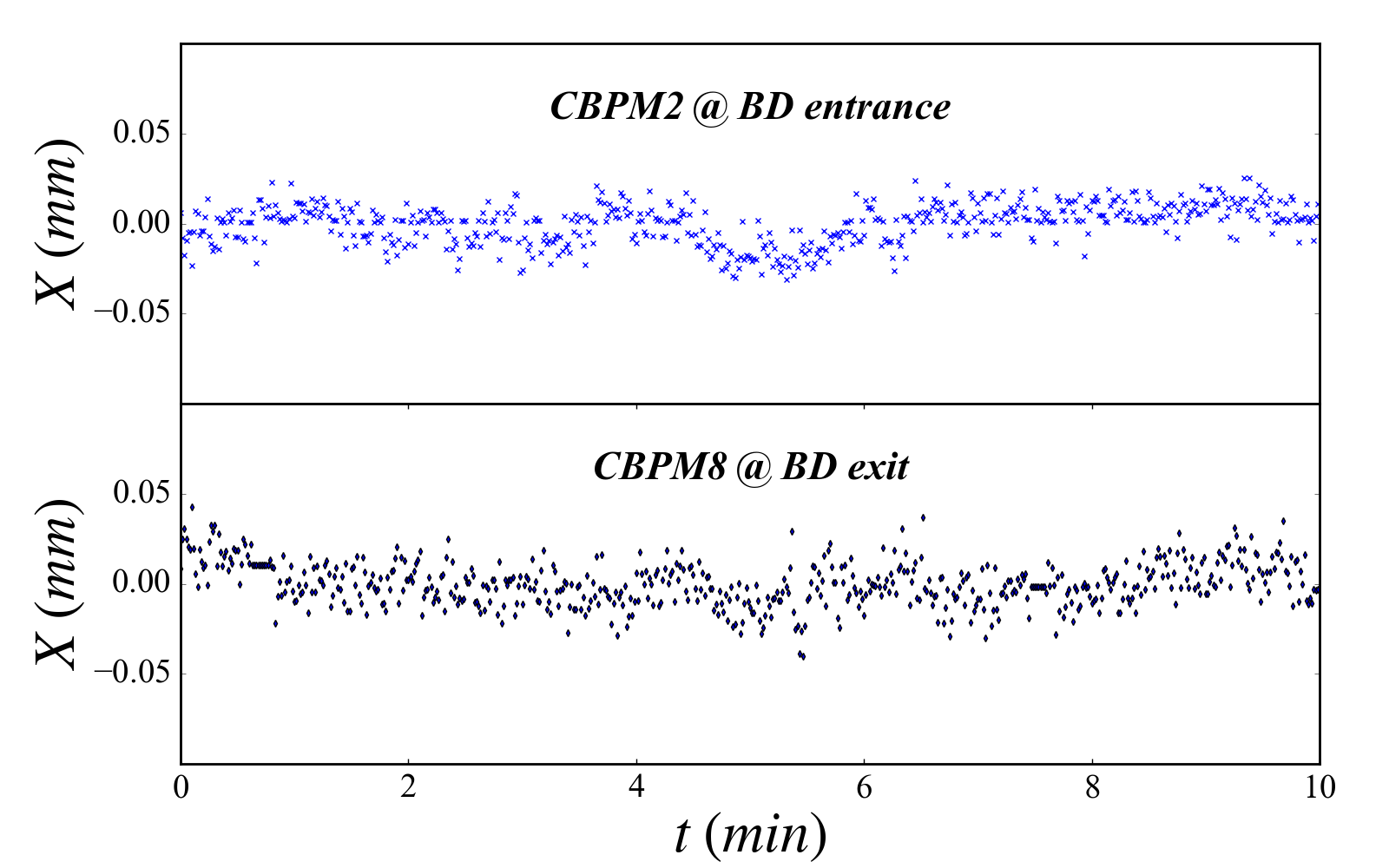}
	\caption{\small{A comparison of the transverse beam central position jitter between the entrance and exit of the BD section. Only imperceptible micro-bunching structure is observed.}}
	\label{fig:jitter-measured}
\end{figure}

\begin{figure}[!htb]
	\centering
	\includegraphics*[width=1.0\linewidth]{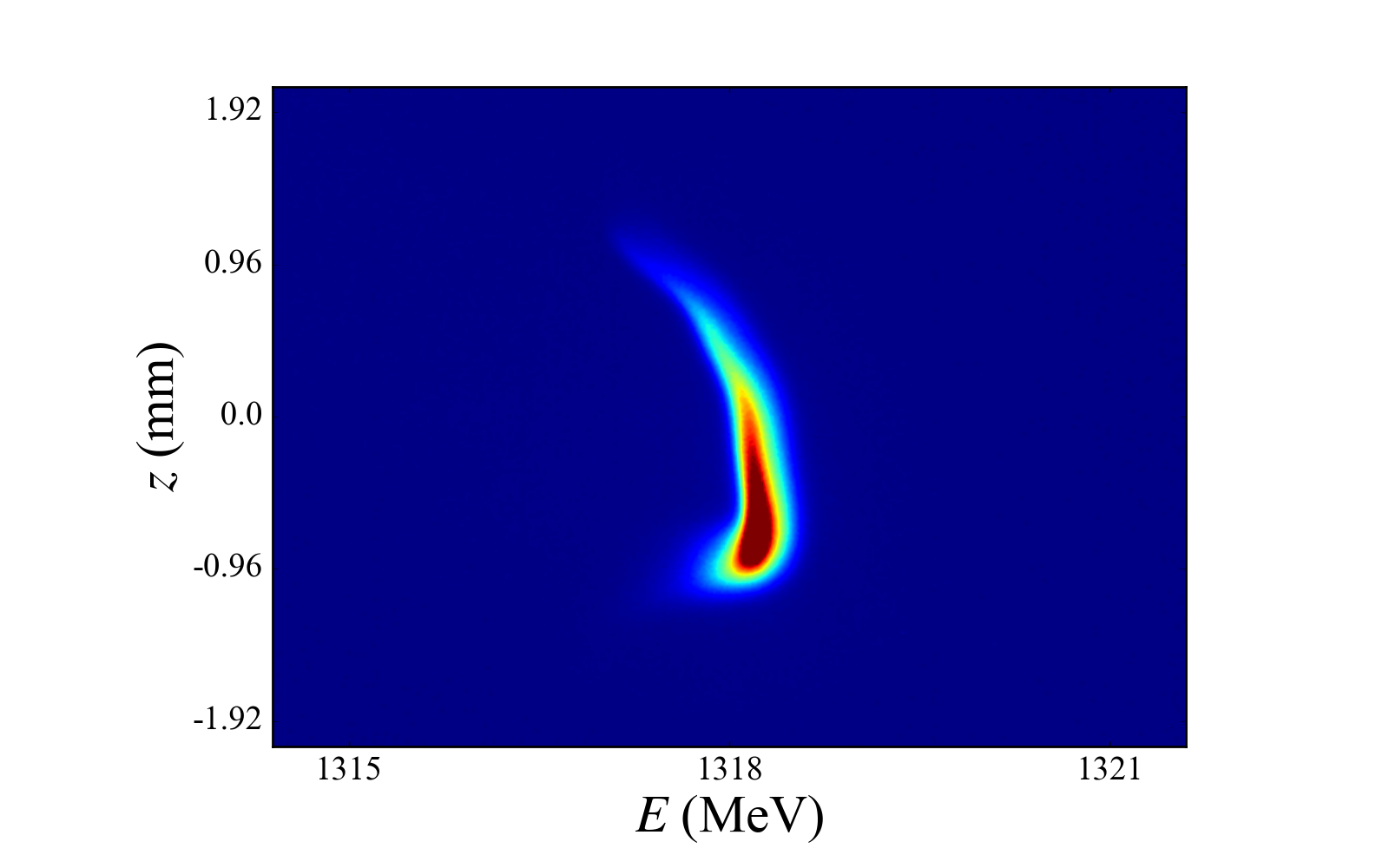}
	\caption{\small{Longitudinal phase space distribution of the electron beam after BD section.}}
	\label{fig:long-ps-measured}
\end{figure}

A comparison of the transverse position jitter before and after the dog-leg is shown in Fig. \ref{fig:jitter-measured}. As the well cancelled dispersion and suppressed CSR effect, the growth of transverse position jitter after the dog-leg is less than \SI{10}{\percent} as is required. The longitudinal phase space at the end of the seeding-FEL line is shown in Fig. \ref{fig:long-ps-measured}. The longitudinal phase space structure is well preserved with negligible micro-bunching growth observed.

\section{Summary}

A beam distribution system of SXFEL-UF is designed to perform a bunch-by-bunch separation of the \SI{50}{\hertz} electron beam to the two undulator lines respectively. With properly optimized optics, some beam collective effects that may spoil the beam quality can be well suppressed. The beam distribution has been installed in the undulator tunnel and started commissioning in Nov. 2021. The commissioning results show that the beam quality after passing through the beam distribution system is well preserved with the present design. The electron beam passes through the beam distribution has been delivered to the downstream FEL line and an exponential FEL power growth in \SI{3}{\nano\meter} has been obtained, which indicates the beam quality after the beam distribution system well satisfies the requirement of soft X-ray FEL. And beyond that, it provides an important reference to the beam distribution system of the future hard X-ray free electron laser facility.

\section{Acknowledgment}

The author would like to thank all the colleagues working on the SXFEL-UF. Special thanks to Duan Gu and Zhen Wang for providing the electron distributions used for this study. Many useful discussions with Bart Fattz and other members of the physics and commissioning team are also gratefully acknowledged.


\begin{thebibliography}{99} 
	\bibitem{Zhao2010}
	Z.T. Zhao. \textit{Storage Ring Light Sources}. Reviews of Accelerator Science and Technology. Vol. 03, No. 01, pp. 57-76 (2010). \url{https://doi.org/10.1142/S1793626810000361}
	
	\bibitem{Huang2021}
	Huang N., et al., \textit{Features and Futures of X-ray Free-Electron Lasers}. The Innovation. 2(2), 100097. \url{https://doi.org/10.1016/j.xinn.2021.100097}
	
	\bibitem{Emma2010}
	Emma, P., Akre, R., Arthur, J. et al. \textit{First lasing and operation of an ångstrom-wavelength free-electron laser}. Nature Photon 4, 641–647 (2010). \url{https://doi.org/10.1038/nphoton.2010.176}
	
	\bibitem{Ishikawa2012}
	Ishikawa, T., Aoyagi, H., Asaka, T. et al. \textit{A compact X-ray free-electron laser emitting in the sub-ångström region}. Nature Photon 6, 540–544 (2012). \url{https://doi.org/10.1038/nphoton.2012.141}
	
	\bibitem{Kang2017}			
	Kang, HS., Min, CK., Heo, H. et al. \textit{Hard X-ray free-electron laser with femtosecond-scale timing jitter}. Nature Photon 11, 708–713 (2017). \url{https://doi.org/10.1038/s41566-017-0029-8}
	
	\bibitem{Prat2020}
	Prat, E., Abela, R., Aiba, M. et al. \textit{A compact and cost-effective hard X-ray free-electron laser driven by a high-brightness and low-energy electron beam}. Nature Photonics 14, 748–754 (2020). \url{https://doi.org/10.1038/s41566-020-00712-8}
	
	\bibitem{Decking2020}
	Decking, W., Abeghyan, S., Abramian, P. et al. \textit{A MHz-repetition-rate hard X-ray free-electron laser driven by a superconducting linear accelerator}. Nature Photonics 14, 391–397 (2020). \url{https://doi.org/10.1038/s41566-020-0607-z}
	
	\bibitem{Ackermann2007}
	Ackermann, W., Asova, G., Ayvazyan, V. et al. \textit{Operation of a free-electron laser from the extreme ultraviolet to the water window}. Nature Photon 1, 336–342 (2007). \url{https://doi.org/10.1038/nphoton.2007.76}
	
	\bibitem{Allaria2012}
	Allaria, E., Appio, R., Badano, L. et al. \textit{Highly coherent and stable pulses from the FERMI seeded free-electron laser in the extreme ultraviolet}. Nature Photon 6, 699–704 (2012). \url{https://doi.org/10.1038/nphoton.2012.233}
	
	\bibitem{Zhao2017}
	Z.T. Zhao, et al., \textit{SXFEL: A Soft X-ray Free Electron Laser in China}. Synchrotron Radiation News, Vol.30(2017) 6, pp.29-33. \url{https://doi.org/10.1080/08940886.2017.1386997} 
	
	\bibitem{Liu2022}
	Liu, B.; Feng, C.; Gu, D.; Gao, F.; Deng, H.; Zhang, M.; Sun, S.; Chen, S.; Zhang, W.; Fang, W.; et al. \textit{The SXFEL Upgrade: From Test Facility to User Facility}. Appl. Sci. 2022,12,176. \url{https://doi.org/10.3390/app12010176}
	
	\bibitem{Zhao2017shine}
	Z.T. Zhao, D. Wang, Z.H. Yang, and L. Yin, “SCLF: An 8-GeV CW SCRF Linac-Based X-Ray FEL Facility in Shanghai”, in Proc. 38th Int. Free Electron Laser Conf. (FEL'17), Santa Fe, NM, USA, Aug. 2017, paper MOP055, pp. 182-184, ISBN: 978-3-95450-179-3, \url{https://doi.org/10.18429/JACoW-FEL2017-MOP055}, 2018.
	
	\bibitem{Hara2016}
	Toru Hara, et al., \textit{Pulse-by-pulse multi-beam-line operation for x-ray free-electron lasers}. Phys. Rev. Accel. Beams 19, 020703 (2016). \url{http://dx.doi.org/10.1103/PhysRevAccelBeams.19.020703}
	
	\bibitem{Hara2018}
	Toru Hara, et al., \textit{High peak current operation of x-ray free-electron laser multiple beam lines by suppressing coherent synchrotron radiation effects}. Phys. Rev. Accel. Beams 21, 040701 (2018). \url{https://doi.org/10.1103/PhysRevAccelBeams.21.040701}
	
	\bibitem{Beukers2019}
	T. Beukers, et al., \textit{A Beam Spreader System for LCLS-II}. SLAC-PUB-17468, \url{https://www.slac.stanford.edu/pubs/slacpubs/17250/slac-pub-17468.pdf}
	
	\bibitem{Balandin2011}
	V. Balandin, et al., \textit{Optics for the Beam Switchyard at the European XFEL}. in Proceedings of IPAC2011, San Sebastián, Spain, WEPC008. \url{https://accelconf.web.cern.ch/IPAC2011/papers/wepc008.pdf}
	
	\bibitem{Milas2012}
	N. Milas, et al., \textit{Switchyard Design: ATHOS}, in Proceedings of FEL2012, Nara, Japan, MOPD37. \url{https://accelconf.web.cern.ch/FEL2012/papers/mopd37.pdf}
	
	\bibitem{DiMitri2013}
	S. Di Mitri, et al., \textit{Cancellation of Coherent Synchrotron Radiation Kicks with Optics Balance}. Phys. Rev. Lett. 110, 014801 (2013). \url{https://doi.org/10.1103/PhysRevLett.110.014801}
	
	\bibitem{elegant}
	M. Borland, \textit{elegant: A Flexible SDDS-Compliant Code for Accelerator Simulation},
	Advanced Photon Source LS-287, September 2000.
	
\end{thebibliography}
\end{document}